\documentclass[twocolumn,prl,showpacs,floatfix]{revtex4}

\usepackage{graphicx}
\usepackage{epsfig}
\usepackage{rotating}
\usepackage{amsmath}
\usepackage{amsfonts}
\usepackage{amssymb}
\usepackage{enumerate}
\usepackage{longtable}
\usepackage{dcolumn}
\usepackage{bm}

\begin{document}

\title{Early breakdown of area-law entanglement at the many-body delocalization transition}

\author{Trithep Devakul$^1$ and Rajiv R. P. Singh$^2$}
\affiliation{$^1$Department of Physics, Princeton University, NJ 08544, USA \\
$^2$Department of Physics, University of California Davis, CA 95616, USA}

\date{\rm\today}

\begin{abstract}
    We introduce the numerical linked cluster (NLC) expansion as a controlled numerical tool for the study of the many-body localization (MBL) transition
    in a disordered system with continuous non-perturbative disorder. 
    Our approach works directly in the thermodynamic limit, in any spatial dimension, and does not rely on any finite size scaling procedure.
    We study the onset of many-body delocalization through the breakdown of area-law entanglement in a generic many-body eigenstate.
    By looking for initial signs of an instability of the localized phase, we obtain a value for the critical disorder, which we believe should be a lower bound for the true value, that is higher than current best estimates from finite size studies.
    This implies that most current methods tend to overestimate the extent of the localized phase due to finite size effects making the localized phase appear stable at small length scales.
    We also study the mobility edge in these systems as a function of energy density, and find that our conclusion is the same at all examined energies.
\end{abstract}

\maketitle

\emph{Introduction}---
The eigenstate thermalization hypothesis (ETH) is a powerful statement relating observables of the high energy eigenstates of a quantum many-body system with 
their thermal expectation values~\cite{eth,rigol-eth1}.
However, this principle can be violated in certain systems with strong enough disorder, where even the high energy eigenstates possess only local entanglement~\cite{mbl-review,rigol-eth2}.  
Anderson localization is a one-body example of this.
An area of key interest is how far this localization persists in a many-body state in presence of interactions~\cite{mbl}.
At what point are interactions strong enough that the localization is destroyed and the system obeys ETH\@? 
This is the problem of many-body localization (MBL) transition, which is a topic of active research both theoretically and experimentally~\cite{vosk-review,muller-review,laumann-sem,vidal-tn,abanin-iom,muller-dynamics,vishwanath-loc,abanin-periodic,nayak-qc,rigol-per,demler-probe,scardicchio-edge,mucciolo-stats,abanin,huse-quasiper,eisert-mbl,altman-random,sondhi-mbl-top,grover-liq,sondhi-order,liangsheng,demler,gross-heis,demarco-hub,bloch,bloch2}.

The surge of interest in many-body localized systems has motivated many numerical studies.
Most studies have focused on exact diagonalization or Lanczos methods which are able to address both sides of the transition in small systems~\cite{abanin-mbl,alet-mbl,huse-rfheisenberg,scardicchio-mbl,pollmann-isingmbl,bardarson-mbl,reichman-mbl,bhatt-mbl,abanin-per,santos-dyn,moore-mbl,moore-mi,rigol-fluc,moore-growth}.  
However, since much about this phase transition is still not well understood, extension of finite size results to the thermodynamic limit can prove difficult.
We would like to examine this phase transition using expansion methods, which provide an alternate way of addressing the thermodynamic limit.
While standard perturbative series expansions are very powerful\cite{oitmaa,series,imbrie}, they suffer from small energy denominators in models with continuous non-perturbative disorder.
Thus, we turn to the numerical linked cluster (NLC) expansion~\cite{nlc,nlc-ent,rigol-nlce}, which does not suffer from this problem of small energy denominators.


In this paper, we provide evidence that for a prototypical model of MBL, approaching the critical disorder from the localized side, the localized phase actually becomes unstable only at increasingly long length scales inaccessible to most numerical techniques.
This implies that finite size numerical studies on the MBL transition tend to overestimate the extent of the localized phase.

\emph{Model}---
The system we study explicitly is the spin-1/2 Heisenberg Hamiltonian with random fields along the $z$ direction,
\begin{equation}
    \mathcal{H} = \sum_i h_i S^z_i + \sum_{\langle i,j \rangle} \vec{S}_i \cdot \vec{S}_j,
    \label{hamil}
\end{equation}
on the 1d chain, where the sum $\langle i,j\rangle$ is over adjacent pairs.
The random field is picked from a uniform distribution $h_i \in [-h,h]$. 
This is one of the simplest models to study many-body localization on, and has been studied numerically in great detail~\cite{huse-rfheisenberg,alet-mbl,abanin-mbl,nayak,scardicchio-mbl,huse-entspread}.
At low $h$, the system is in a thermalizing phase obeying the ETH\@, while
at high $h$, the system is in the localized (MBL) phase.

To identify these different phases, we focus on the entanglement properties of the eigenstates.  
The typical measure for entanglement in a pure state bipartitioned into two parts $A$ and $B$ is the von Neumann entropy, defined for some state $\left|{\Psi}\right>$ as
\begin{equation}
    s(\left|{\Psi}\right>) = -\text{Tr}(\rho_A \ln \rho_A,)
    \label{vnent}
\end{equation}
where $\rho_A = \text{Tr}_B\left|{\Psi}\right>\left<{\Psi}\right|$ is the reduced density matrix, obtained by tracing over all external degrees of freedom from the density matrix.

A typical eigenstate in an ETH obeying system will exhibit thermal volume law entanglement.
The entanglement entropy will approach the classical thermal entropy (required by ETH) and scale with the volume of the regions $A$ and $B$.
In the localized phase, the eigenstates will instead obey an area-law, scaling with the area between $A$ and $B$.  
This can be understood by regarding them as simultaneous eigenstates of many local operators~\cite{mbl-review}: only due to mixing 
contained in operators near the boundary will one get contributions to the entanglement, which therefore grows with the area of the bipartitioning.

\emph{Numerical Linked Cluster Expansions}---
NLC is similar to perturbative series expansions in that interactions within clusters of increasing sizes must be considered, but rather than perturbatively treating each interaction within a cluster, we solve them numerically, typically by exact diagonalization.  
Our treatment of disorder in NLC is different from usual~\cite{spin-glass,rigol-nlce}, allowing us to deal with continuous non-perturbative disorder.
The procedure is outlined briefly here.

Let $N$ be the order to which we wish to do the calculation.
The order of the calculation defined as the number of spins in the largest cluster considered.
We identify a finite size region of the infinite system to work with.
For the chain, this is simply a $2(N-1)$ length chain, with a bipartitioning cut in the middle.
Each of the sites $i$ is assigned a field $h_i$, which is held fixed until the calculation is complete.
The reason for this choice of system size is so that the results remain correct for the infinite system, to the desired order, as explained later.

We define a cluster $c$ to be a set of sites.
A Hamiltonian $\mathcal{H}_c$ can be obtained considering only the spins in $c$, and diagonalized numerically to obtain the eigenstates $\{\left|{\alpha^c}\right>\}$ with eigenvalues $\{\epsilon_\alpha^c\}$ labeled by $\alpha$.
Our quantity of interest, the eigenstate averaged entanglement entropy, can then be calculated as
\begin{equation}
    S(c) = \sum_{\alpha} \frac{e^{-\beta \epsilon_\alpha^c}}{\mathcal{Z}}s(\left|{\alpha^c}\right>),
    \label{sdef}
\end{equation}
where $\mathcal{Z} = \sum_\alpha e^{-\beta \epsilon_\alpha^c}$ is the normalization factor and $\beta=1/T$ is the inverse temperature.

The entropy for the infinite lattice $\mathcal{L}$ can be expressed as a sum over the \emph{weight} $\widetilde{S}(c)$ of all clusters $c$ that can be embedded in the lattice:
$S(\mathcal{L}) = \sum_{c } \widetilde{S}(c)$.
The weight of a cluster is then defined recursively by the principle of inclusion and exclusion:\cite{nlc}
\begin{equation}
    \widetilde{S}(c) = S(c) - \sum_{c' \subset c} \widetilde{S}(c').
\end{equation}

One can show that only connected clusters which cross the boundary can have a nonzero weight.
First, if a cluster does not cross the boundary there can obviously be no entanglement in it or its subclusters, so the weight is trivially zero.
Second, proving that only connected clusters can contribute simply amounts to proving that $S$ obeys the \emph{linked cluster property}, that is, for a cluster with two disconnected components $c_1$ and $c_2$, $S(c_1 \cup c_2) = S(c_1) + S(c_2)$.  
This follows from the fact that $\mathcal{H}_{c_1 \cup c_2} = \mathcal{H}_{c_1} \oplus \mathcal{H}_{c_2}$.
Thus, we must simply consider connected clusters of up to size $N$ that have sites on both sides of the partition.  
Our finite size representation was chosen to contain all the necessary clusters of the infinite system up to order $N$.
The count for the clusters crossing the boundary scales with the area thus guaranteeing an area-law \emph{as long as NLC converges}.

Using this, we can obtain a series $a_n$ whose sum gives the total eigenstate averaged entanglement entropy per unit area $S_\text{area}$,
\begin{equation}
    S_\text{area} = \frac{1}{L_\text{cut}}\sum_{n=0}^N a_n \qquad ; \qquad a_n = \sum_{c,|c|=n} \widetilde{S}(c),
    \label{andef}
\end{equation}
where $L_\text{cut}$ is simply $1$ for the chain.
Finally, \emph{the entire calculation must be repeated for different realizations of $\{h_i\}$ to obtain a disorder averaged value for $a_n$.}~\cite{footnote1}

The NLC scheme used here is slightly different from what has typically been used for random systems in the past, where different embeddings of the same graph are treated identically and one does not need a consistent finite system~\cite{spin-glass,rigol1,rigol2}.
The more standard scheme has been applied to study MBL systems with discrete disorder~\cite{rigol-nlce}, which allows one to perform disorder averaging before subgraph subtraction.
When there is continuous disorder, partial disorder averaging over a finite number of realizations means that the linked cluster property is only approximate, and thus large errors will build up at high orders.
In our approach, the linked cluster property is guaranteed and one is free to average over many realizations of the system.
This is a key aspect of our calculation which allows us to treat disordered systems with continuous non-perturbative randomness.

\emph{Does it converge?}---
We first examine $T=\infty$.
If entanglement satisfies a thermal volume-law, interpreting $n$ as a proxy length scale~\cite{nlc-ent}, we expect $a_n$ to eventually saturate to the volume-law constant $\ln(2)/2$ at high enough $n$~\cite{footnote2}.
We should note that our model (Eq~\ref{hamil}) does not possess a strongly thermalizing regime, due to the integrability at $h=0$, and therefore we do not yet see this saturation to the thermal value within our range of $n$~\cite{footnote3}.
In the localized phase, the additional entanglement due to the addition of one site far away from the cut should become exponentially small with distance, so we expect $a_n$ to decay exponentially to 0 once $n$ is larger than some localization length $\xi$.
We \emph{define} the MBL phase in our study to be one in which the sum of $a_n$ converges exponentially.

\begin{figure}
    \centering
    \includegraphics[width=0.5\textwidth]{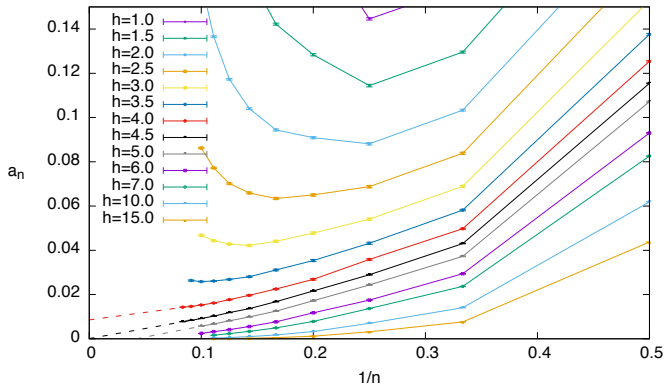}
    \caption{The $n$th order area-law contribution $a_n$ (Eq.~\ref{andef}) at $T=\infty$.
Near $h_c$, data has been averaged over more than $3\times 10^5$ disorder realizations of the chain, and error bars show the standard error of the mean.
Dashed lines are a demonstration of the linear extrapolation to $a_\infty$ by fitting the last 4 terms in the series.
}
\label{fig:an}
\end{figure}

Fig~\ref{fig:an} shows $a_n$ for a range of $h$ values.  
To estimate convergence or divergence, a linear extrapolation to $1/n=0$ can be performed.
If the extrapolation predicts $a_\infty \geq 0$, we argue that this corresponds to a breakdown of area-law.
Although we expect $a_n$ to eventually go to 0 exponentially in the localized phase, this would only happen when our cluster sizes are much greater than the localization length scale.
This can be more clearly shown by examining the ratio of terms, which we will discuss next.
Note that the case of an area-law with logarithmic corrections would correspond to one where $a_n$ heads linearly to $0$, which we consider in this analysis the boundary between convergence and divergence.

\begin{figure}
    \centering
    \includegraphics[width=0.5\textwidth]{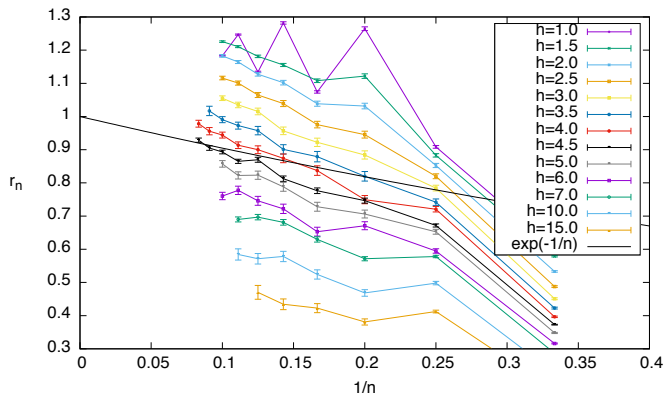}
    \caption{Plot of the ratios $r_n = a_n/a_{n-1}$ versus $1/n$.
    Also shown is the line $\exp(-1/n)$, above which we argue the expansion cannot converge exponentially.
    }
\label{fig:rn}
\end{figure}

Let us define the ratio of the (disorder averaged) series terms $r_n = a_n/a_{n-1}$.
In the MBL phase, we can say more about the overall trend of $r_n$.
Again interpreting $n$ as a proxy length scale, for large $n$, we expect $a_n$ to decrease exponentially with potentially power-law prefactors.  
The leading contributions at $n \gg \xi$ should be of the form $a_n = C n^{-k}\exp(-n/\xi)$, where $k$ is some positive number, $\xi$ is the localization length, and $C$ is some arbitrary constant~\cite{mbl-review}.
Therefore, in the large $n$ limit, discarding terms smaller than $1/n$, we expect
\begin{equation}
    r_n = a_n/a_{n-1} = (1-k/n) \exp (-1/\xi).
    \label{rndef}
\end{equation}
Therefore, plotting $r_n$ versus $1/n$, $r_n$ should approach $r_\infty=\exp(-1/\xi)$ from below, with a slope of $-k$.
However, near the transition $\xi$ can become very large and we do not actually see this behavior within our range of attainable $n$.
We can, however, predict whether this kind of exponential convergence is possible given the behavior of the series at finite $n$.

Fig~\ref{fig:rn} shows the behavior of $r_n$ for our range of $h$ and $n$.
The trend seems to be for $r_n$ to increase steadily with $n$ (although eventually $r_n$ must approach 1 in the delocalized phase).
If the system is in the MBL phase and the series is to converge exponentially, we expect that at some $n \gg \xi$, $a_n$ will begin decreasing exponentially and $r_n$ will begin heading towards $r_\infty$ with slope $-k$.
Barring bizarre behavior such as $r_n$ increasing to some high value and then suddenly decreasing before finally increasing again towards $r_\infty$, this places a restriction on what $r_n$ can be when $a_n$ begins its exponential decay.
Because the slope of the approach is negative or 0, $r_n \leq r_\infty$.  
But also $n \gg \xi$, which means this decay can only begin occurring when
\begin{equation}
    r_n \leq r_\infty = \exp(-1/\xi) < \exp(-1/n).
    \label{rnineq}
\end{equation}

Therefore, once $r_n$ has increased above $\exp(-1/n)$, $a_n$ cannot converge exponentially.
This is not a rigorous claim, but should be valid as long as $a_n$ behaves in a regular manner.
This clearly shows (in Fig~\ref{fig:rn}) that the series for $h=4.0$  cannot converge exponentially, and thus is not in the MBL phase.
However, the series for $h>4.5$ are still within this region, and thus may diverge or converge.  
Hence, our result should serve as a lower bound, with our best estimate being at $h_c = 4.5\pm0.1$.  
Going to higher order in NLC can further refine this value.  

\emph{Discussion}---
A way of viewing our result~\cite{huse-thanks} is that we are seeing an instability to thermalization of an \emph{almost-localized} regime~\cite{nandkishore-nearly-localized}.  
That is (in Fig~\ref{fig:an}), initially $a_n$ acts quite localized in that it is much smaller than the thermal value and is getting smaller as $n$ is increased, but may start increasing at higher $n$, signaling the ``onset'' of thermalization.
This onset of thermalization moves to higher $n$ as $h$ is increased, but goes beyond our range of accessible $n$ after $h=3.5$.  
However, by looking in a sensitive way for initial signs of an instability, we are able to place a lower bound for $h_c$ at $4.5\pm0.1$.

To understand how our analysis is more sensitive to this transition than other methods, let us focus on the entanglement per unit volume $S_\text{vol}$.  
$S_\text{vol}$ decreasing with system size is often associated with area-law entanglement and therefore localization~\cite{alet-mbl}.
In our study, $S_\text{vol}$ for a system of size $N$ would correspond to the quantity $S_\text{vol}=(1/N)\sum_{n=0}^N a_n$.
So even if $a_n$ had already ``turned up'' and was increasing (clearly thermalizing), $S_\text{vol}$ would not begin increasing until $a_n$ had increased above the mean of all the previous terms in the series.
Our analysis predicts this upturn, which itself would precede estimates from finite size systems using $S_\text{vol}$.

Other methods, which are more focused on seeing the full onset of thermalization, do not observe the transition near our bound~\cite{huse-rfheisenberg,alet-mbl,abanin-mbl,scardicchio-mbl}.  
Results from Lanczos on systems of up to $22$ sites with finite size scaling show evidence for a transition at $h_c \approx 3.7$~\cite{alet-mbl}.
However, the scaling exponent $\xi\sim {(h-h_c)}^{-\nu}$ obtained from finite size scaling strongly violates the Harris-Chayes bound in 1d~\cite{chayesharris,anushya-fss}, evidence that perhaps they are still far from the true critical point.
This implies that finite size effects are significant, even in the systems accessible to Lanczos, and some corrections to the finite size scaling is needed.
These effects cause an overestimate of the stability of the MBL phase, which actually becomes unstable to thermalization earlier only at much longer length scales.
Note that an interesting alternate possibility is the existence of an intermediate phase between the ETH and MBL phase, with neither thermal nor area-law entanglement~\cite{grover,scardicchio-nonerg-extended,xiaopeng}, which we do not pursue further.

This onset of thermalization at high order is what one would expect from a long lengthscale delocalization mechanism.
In studies of this transition by a renormalization group approach, one also finds that the transition is driven by rare metallic inclusions~\cite{potter,vosk}.
Near the transition, these are rare enough that small systems look localized, but actually become thermalizing at long length scales. 


\begin{figure}[t]
    \centering
    \includegraphics[width=0.5\textwidth]{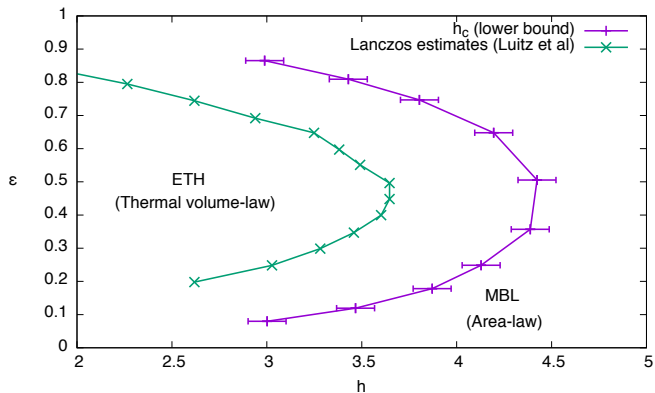}
    \caption{The phase diagram with (scaled) energy and disorder on the axes, identified by interpolation of the intercept method (Fig~\ref{fig:an}) up to order 10.  
        Error bars represent confidence in our interpolation.
        Also shown is the $h_c$ obtained by finite size scaling of Lanczos results from Ref~\cite{alet-mbl} (error bars not shown).
    }
\label{fig:phasediagram}
\end{figure}

\emph{The mobility edge}---
Finally, we can observe the transition at different energy windows by varying $\beta=1/T$ in Eq.~\ref{sdef} of our NLC calculation, thus probing states at a given energy defined
by the thermal ensemble.
Following the same arguments as at $T=\infty$, we can obtain estimates for $h_c$ at a given temperature or energy density.
Fig~\ref{fig:phasediagram} shows our estimates for various $\beta$ values, with the scaled energy $\epsilon$ on the vertical axis: $\epsilon = (E-E_\text{min})/(E_\text{max}-E_\text{min})$, where $E_\text{min}$ and $E_\text{max}$ are the lowest and highest energies in the energy spectrum.
The shape of our estimates are very similar to those obtained in previous numerical calculations~\cite{alet-mbl,abanin-mbl}, along with the slight asymmetry expected around $\epsilon=1/2$~\cite{laumann-edge}.
As with the case at $T=\infty$, we find our estimates are consistently higher than previous numerical calculations.

There is much debate on whether a mobility edge exists in the thermodynamic limit.  
Numerical results suggest the presence of such an edge~\cite{alet-mbl,abanin-mbl,laumann-edge}, but there are also arguments against it~\cite{schuilaz-bubbles}.
While our phase diagram shows a similar shape as previous numerical studies, our analysis gives a lower bound for $h_c$, and hence does not negate the claims of the absence of a mobility edge.
If the transition actually occurs at a single $h_c$ for all energies,
the fact that our estimates are lower away from the center of the spectrum would indicate that much larger length scales would be needed to observe delocalization and that finite size effects would be much stronger in those regions.

\emph{Conclusions}---
In conclusion, we have studied the MBL transition in the random field Heisenberg model using NLC expansions.  
We focus on the breakdown of the area-law of entanglement in the eigenstates of the Hamiltonian.
Our approach works directly in the thermodynamic limit and does not rely on any finite size scaling.
By looking for signs of instability in the MBL phase, we are able to estimate a lower bound for the critical disorder in the thermodynamic limit.
At all energies examined, our $h_c$ estimates are consistently higher than found by finite size studies.
This implies that numerical methods which look for the full onset of thermalization tend to overestimate the extent of the MBL phase, which actually becomes unstable earlier but only at much longer length scales.
Near the transition, finite size effects are significant and hence caution must be taken when relating to the infinite system.

Our result can be readily verified by cold atoms experiments, which are able to present very well characterized systems~\cite{gross-heis, demarco-hub, bloch, bloch2}.
If the 1d random field Heisenberg model is experimentally realized, measurements of the critical disorder for large systems should lie above our estimate.
We may also extend our result for other similar models of the disorder driven MBL transition, where delocalization also occurs over a long lengthscale~\cite{potter,vosk}, and suggest that the true critical point would be higher than finite size scaling estimates from small systems.
Also of interest are quantum chaotic Wannier-Stark systems~\cite{wannier1,wannier2}, which are experimentally accessible and possess a localization-delocalization transition in the absence of disorder.

\begin{acknowledgments}
We would like to thank David Huse for many valuable discussions. This work is supported in part by NSF grant number DMR-1306048.
\end{acknowledgments}


\begin{thebibliography}{100}
\bibitem{footnote2} This was confirmed by performing the expansion for random pure states.
\bibitem{footnote3} In a model with a strongly thermalizing phase, such as the Floquet model studied in~\cite{liangsheng}, we do see $a_n$ saturating at the thermal value.
\end{thebibliography}
\end{document}